# An Enhanced Semidefinite Relaxation Model Combined with Clique Graph Merging Strategy for Efficient AC Optimal Power Flow Solution

Zhaojun Ruan, and Libao Shi, *Senior Member, IEEE*

*Abstract*—Semidefinite programming (SDP) is widely acknowledged as one of the most effective methods for deriving the tightest lower bounds of the optimal power flow (OPF) problems. In this paper, an enhanced semidefinite relaxation model that integrates tighter λ-based quadratic convex relaxation, valid inequalities, and optimality-based bound tightening algorithms derived in accordance with the branch thermal limit boundary surface into the SDP framework is presented to further tighten the lower bounds of the feasible region of OPF problems, effectively combining the advantages of these recent advancements. Additionally, the utilization of chordal decomposition in the complex matrix formulation of SDP can significantly accelerate the solution time. Notably, for the same SDP problem, different chordal decompositions can result in varying solution time. To address this problem, this paper proposes a clique graph merging strategy within the complex matrix SDP framework, which assesses clique sizes and the computational burden on interior-point solvers, as well as reducing the need for hyperparameter tuning and further enhancing the solution efficiency. Finally, the proposed hybrid relaxation model is evaluated using MATPOWER and PGLib-OPF test cases, demonstrating its effectiveness in reducing the optimality gap and validating its computational performance on test cases with up to 13659-node.

## I. INTRODUCTION

IT is known that power flow analysis is the foundation of power system research, addressing critical challenges related to the planning, operation, and control of power systems. In order to further optimize the system operation behavior and enhance the economic efficiency of power system operation, Carpentier introduced the OPF problem in 1962 [1]. The OPF aims to achieve the optimal operation of power grid with a certain goal, such as minimizing system generation cost, reducing power loss, or refining voltage profile, while complying with a set of physical and operational constraints, such as Ohm's Law, Kirchhoff's Laws, branch thermal limits, and system stability requirements. Up to now, how to efficiently and quickly solve large-scale OPF problems remains significant challenges. In particular, the AC power flow equations pose considerable obstacle to the solution of OPF problem owing to their highly nonlinear and non-convex nature, which encapsulates the complex interdependence between power and voltage. Theoretically, finding the global optimal solution of the AC OPF problem is a typical NP-hard problem [2], [3].

In response to these challenges, over the past decade, a lot of significant studies have focused on convex relaxation techniques to solve the OPF problems, mainly involving second-order cone relaxation (SOCR) [4], [5], [6] quadratic convex relaxation (QCR) [7], [8], [9] semidefinite relaxation (SDR) [10], [11], [12], and hierarchies of moment or sum-of-squares relaxations [13], [14], [15]. Among these, the SDR has proven to be the most stringent relaxation technique for the power flow equations. It is worth noting that recent studies have demonstrated that the SDR technique can achieve global optimal solutions in certain practical cases [16]. To enhance the tightness of the relaxation, Molzahn and Josz et al. [15], [17] proposed using moment information to construct more refined SDR, at the expense of increasing computational complexity to improve accuracy. However, the inclusion of the high-order moment information renders this technique less feasible for large-scale power system applications.

Additionally, recent studies have suggested that the QCR technique, based on recursive McCormick envelopes, encounters difficulties in capturing the convex hull of trilinear terms [18]. Sundar et al. [19] advocated combining tighter λ-based QCR with optimality-based bound tightening (OBBT) techniques, which can reduce the optimality gap and provide tighter lower bounds than SDR in some cases. Nonetheless, owing to the high computational burden of the OBBT algorithm, it requires significant computation time even under ideal parallel conditions, making it challenging to apply to systems with more than 1000 buses [20]. Additionally, a hybrid approach [21] was proposed to enhance the accuracy of SDR by integrating QCR and SDR techniques, introducing new valid inequalities and boundary tightening strategies. Unfortunately, the computational inefficiency of SDR limits its scalability to large-scale cases. Bingane et al. [22], [23] proposed combining semidefinite programming (SDP) and the reformulation-linearization technique to obtain a more computationally efficient approach called strong tight-and-cheap relaxation (STCR), which makes a trade-off between SDR and SOCR for large-scale cases with respect to computation time and optimality gap.

Given the circumstances, constructing chordal relaxation

Z. Ruan and L. Shi are with the Tsinghua Shenzhen International Graduate School, Tsinghua University, Shenzhen 518055, China (e-mail: rzj23@mails.tsinghua.edu.cn; shilb@sz.tsinghua.edu.cn).

(CHR) of the OPF problem by using chordal decomposition techniques [10], [11], [24], [25] provides a promising way to improve the computational efficiency of sparse large-scale SDP problems. Nevertheless, selecting an appropriate chordal extension with controllable computational complexity remains daunting challenges. A well-chosen chordal decomposition technique can accelerate SDP by 10 to 20 times, whereas a poorly selected decomposition technique may still require extensive computation time [10], [26], [27]. Consequently, some scholars tried to solve this problem by introducing clique merging techniques with varying degrees of success [10], [27], [28], [29]. To achieve high-quality chordal decompositions, Molzahn et al. [10] initially proposed a greedy strategy for clique merging to address large-scale OPF problems. On this basis, Sliwak et al. [27] enhanced the solution efficiency through a clique tree merging strategy, although this method involves numerous hyperparameters, and changing these parameters will affect the quality of the clique merging. Additionally, to tackle the issue of clique merging in the context of "nephew-uncle" or "sibling" relationship, Garstka et al. [30] proposed a clique graph merging strategy applied to the alternating direction method of multipliers solvers. However, using computation time as the sole criterion for evaluating clique merging results may lead to the merging of large cliques, which may not necessarily enhance computational efficiency. It is worth noting that some studies have demonstrated that relaxing the non-convex constraints prior to converting from complex numbers to real numbers can improve solution efficiency [31].

Inspired by the existing research, this paper proposes an enhanced SDR model to solve the large-scale OPF problem. By integrating the model intersection technique, valid inequalities, and bound tightening algorithm into the SDP framework, the lower bound of the feasible region of OPF problem can be further tightened. Further, a clique graph merging strategy within the complex matrix SDP framework is proposed to improve the solution efficiency. The corresponding simulations and comparative analysis are performed on different sizes of test cases to demonstrate the effectiveness and efficiency of the proposed model and method.

The rest of the paper is structured as follows. Section II reviews the formulations of the AC OPF problem based on different convex relaxation techniques. Section III introduces three methods for tightening convex relaxations and proposes a hybrid relaxation model. Section IV presents the clique merging strategy to further improve the efficiency of solving OPF problems. Section V demonstrates the advantages of the proposed model and method through various test case studies, and some important findings are outlined in Section VI.

## II. OPTIMAL POWER FLOW OVERVIEW BASED ON CONVEX RELAXATION TECHNIQUE

In this section, firstly, several different relaxation forms for solving OPF problem are systematically reviewed, and then a tighter and more effective OPF-oriented relaxation model is derived.

Regarding a standard $n$-bus power network $\mathcal{G} = (\mathcal{N}, \mathcal{E})$ with $ng$ generators, where $\mathcal{N} = \{1,2,\ldots,n\}$ and $\mathcal{E} \subseteq \mathcal{N} \times \mathcal{N}$ represent the sets of all buses and branches, respectively, including transformers, phase shifters, and shunt compensation devices. Each branch $l \in \mathcal{E}$ has a '*from*' end $k$ (on the tap side) and a '*to*' end $m$, marked as $l = (k, m)$, allowing for parallel branches (i.e., more than one branch between the same terminal buses). Typically, a reference bus $r \in \mathcal{N}$ is also specified. In the following, let $\Re(\cdot)$ and $\Im(\cdot)$ denote the real and imaginary parts, respectively, let $\mathbb{R}, \mathbb{R}_+, \mathbb{C}, \mathbb{H}^n$ and $\mathbb{S}^n$ denote the set of real numbers, non-negative real numbers, complex numbers, n-order Hermitian matrices, and n-order symmetric matrices, respectively, and let the bold **i** represent the imaginary unit of a complex number.

### A. AC Optimal Power Flow

So far, many mathematical models have been used to describe the AC OPF problems [32], [33]. In recent studies, the power flow equations are described by the following two primary models: the bus injection model (BIM) [24], [32] and the branch flow model (BFM) [4]. The models discussed later in this paper will be characterized using the BIM, which captures the physical relationships between current *I*, voltage *V*, power *S*, and admittance *Y*. The specific expressions of the power flow equations described using BIM are as follows:

$$S_i^{gen} - S_j^{load} = \sum_{(i,j)\in\mathcal{E}} S_{ij} + \sum_{(j,i)\in\mathcal{E}} S_{ji} \quad \forall i,j \in \mathcal{N} \quad (2a)$$

where

$$S_{ij} = V_i I_{ij}^* = Y_{ij}^* V_i V_i^* - Y_{ij}^* V_i V_j^* \quad (i,j), (j,i) \in \mathcal{E} \quad (2b)$$

where $S_i^{gen}$ represents the apparent power generated by the generator located at bus $i$. $S_j^{load}$ represents the apparent power of the load located at bus $j$. $V_i, V_j$ represent the voltages of bus $i$ and bus $j$, respectively.

The AC power flow equations described in (1a) and (1b) are the core of the non-convex and nonlinear characteristics of the OPF problem. In addition to the power flow equation constraints, it is also necessary to consider the operational constraints of the real-world power grids [34], mainly including:

- Generator power capacities:

$$\Re(S_i^{gen(l)}) \leq \Re(S_i^{gen}) \leq \Re(S_i^{gen(u)}) \quad \forall i \in \mathcal{N}_{ng} \quad (2c)$$
$$\Im(S_i^{gen(l)}) \leq \Im(S_i^{gen}) \leq \Im(S_i^{gen(u)}) \quad \forall i \in \mathcal{N}_{ng} \quad (2d)$$

where, $S_i^{gen(l)}$ and $S_i^{gen(u)}$ denote the lower and upper bounds of $S_i^{gen}$, respectively.

- Voltage magnitude limits:

$$\left(v_i^{(l)}\right)^2 \leq |V_i|^2 \leq \left(v_i^{(u)}\right)^2 \quad \forall i \in \mathcal{N} \quad (2e)$$

where $|V_i|$ denotes the voltage magnitude of bus $i$, and $\left(v_i^{(l)}, v_i^{(u)}\right) \in (\mathbb{R}_+, \mathbb{R}_+)$ denotes its lower and upper bounds.

- Voltage phase angle difference limits:

$$\tan\left(\theta_{ij}^{(l)}\right)\Re(V_i V_j^*) \leq \Im(V_i V_j^*) \leq \tan\left(\theta_{ij}^{(u)}\right)\Re(V_i V_j^*) \quad \forall (i,j) \in \mathcal{E} \quad (2f)$$

where $\theta_{ij}^{(l)}$ and $\theta_{ij}^{(u)}$ represent the lower and upper bounds of voltage phase angle difference (i.e., $\theta_{ij} = \theta_i - \theta_j$).

- Branch thermal limits:

$$|S_{ij}| \leq s_{ij}^{(u)} \quad \forall (i,j), (j,i) \in \mathcal{E} \tag{2g}$$

where $|S_{ij}|$ and $s_{ij}^{(u)}$ denote the apparent power magnitude and its upper bound, respectively.

- Reference bus constraint:

$$\angle V_r = 0 \tag{2h}$$

where $\angle V_r$ is the voltage phase angle of the reference bus.

For the AC OPF problem, the objective function is typically defined as minimizing generator fuel cost as shown in (1i) or branch loss. For other non-convex objective functions, the linearization techniques can be leveraged to convert them into convex functions.

$$\min \sum_{i \in \mathcal{N}_{ng}} a_i \left( \Re(S_i^{gen}) \right)^2 + b_i \Re(S_i^{gen}) + c_i \tag{2i}$$

It should be noted that the objective function for minimizing branch loss can be regarded as a special case of (1i), where $a_i = 0, b_i = 1, c_i = 0$ ($\forall i \in \mathcal{N}$). Hence, this study will mainly focus on (1i) without loss of generality [21].

Here, by introducing a new variable $W_{ij}$ [11], [12] ($V_i V_j^* = W_{ij} \quad \forall (i,j) \in \mathcal{E}$), this paper presents several typical expressions for AC OPF optimization models based on different convex relaxation techniques, and will be elaborately described in the following sections.

For the AC OPF model, abbreviated as AC-OPF, it is regarded as the original AC OPF model with highly non-convex and nonlinear characteristics. For the sake of simplicity in presentation, the transformers, phase shifters, and shunt reactors are ignored, although these are included in our numerical simulations. The specific AC-OPF model is given as follows.

| **AC-OPF Model** |
|---|
| **minimize:** (1i) |
| **subject to:** (1a), (1c), (1d), (1g), (1h) |
| $\quad S_{ij} = Y_{ij}^* W_{ii} - Y_{ij}^* W_{ij} \quad (i,j) \in \mathcal{E} \quad (2j)$ |
| $\quad S_{ji} = Y_{ij}^* W_{jj} - Y_{ij}^* W_{ij}^* \quad (j,i) \in \mathcal{E} \quad (2k)$ |
| $\quad \left(v_i^{(l)}\right)^2 \leq W_{ii} \leq \left(v_i^{(u)}\right)^2 \quad \forall i \in \mathcal{N} \quad (2l)$ |
| $\quad \tan(\theta_{ij}^{(l)}) \Re(W_{ij}) \leq \Im(W_{ij}) \leq \tan(\theta_{ij}^{(u)}) \Re(W_{ij}) \quad \forall (i,j) \in \mathcal{E} \quad (2m)$ |
| $\quad W_{ii} = |V_i|^2 \quad \forall i \in \mathcal{N} \quad (2n)$ |
| $\quad W_{ij} = V_i V_j^* \quad (i,j) \in \mathcal{E} \quad (2o)$ |

*B. AC OPF Model Based on QCR*

It is known that the QCR is an enhancement of the SOCR technique, and has been successfully applied to the AC OPF problem [9], [35]. In this section, the AC OPF model based on QCR, abbreviated as AC-OPF-QCR, will be addressed carefully. In polar coordinates (i.e., $V_i = v_i \angle \theta_i \quad \forall i \in \mathcal{N}$), equations (1n) and (1o) are rewritten to incorporate more complete information about the voltage variables at both ends of the branch, thereby minimizing the optimality gap. The relationship between the variable $W$ and the voltage magnitude $v$, the phase angle $\theta$ in polar coordinates can be described as:

$$W_{ii} = v_i^2 \quad i \in \mathcal{N} \tag{3a}$$
$$\Re(W_{ij}) = v_i v_j \cos(\theta_{ij}) \quad \forall (i,j) \in \mathcal{E} \tag{3b}$$
$$\Im(W_{ij}) = v_i v_j \sin(\theta_{ij}) \quad \forall (i,j) \in \mathcal{E} \tag{3c}$$

When constructing its convex hull, since the phase angle difference between the voltage at both ends of the branch in actual high-voltage transmission network power flow usually does not exceed $10°$, it can be used as a boundary condition, which is helpful to construct the convex hull of the square voltage term, bilinear product term, and trigonometric term. As discussed in [9], the first two nonlinear terms in the QCR (2b) and (2c) use the well-known McCormick envelopes [36], given by the following T-CONV and M-CONV, respectively:

$$\langle x^2 \rangle^T \equiv \begin{cases} \hat{x} \geq x^2 \\ \hat{x} \leq \left(x^{(u)} + x^{(l)}\right)x - x^{(u)} x^{(l)} \end{cases} \tag{3d}$$

$$\langle xy \rangle^M \equiv \begin{cases} \widehat{xy} \geq x^{(l)} y + y^{(l)} x - x^{(l)} y^{(l)} \\ \widehat{xy} \geq x^{(u)} y + y^{(u)} x - x^{(u)} y^{(u)} \\ \widehat{xy} \leq x^{(l)} y + y^{(u)} x - x^{(l)} y^{(u)} \\ \widehat{xy} \leq x^{(u)} y + y^{(l)} x - x^{(u)} y^{(l)} \end{cases} \tag{3e}$$

where $(x^{(l)}, x^{(u)}), (y^{(l)}, y^{(u)}) \in (\mathbb{R}_+, \mathbb{R}_+)$ denote the variable bounds of $x$ and y.

Assuming that the difference of phase angle $\theta_{ij}$ satisfies $-\pi/2 \leq \theta_{ij}^{(l)} \leq \theta_{ij}^{(u)} \leq \pi/2$, then the cosine and sine functions of $\theta_{ij}$ are given by the following expressions:

$$\langle \cos(\theta_{ij}) \rangle^C \equiv \begin{cases} \widehat{\theta^c} \leq 1 - \frac{1-\cos(\theta_{ij}^{(m)})}{\left(\theta_{ij}^{(m)}\right)^2} \theta_{ij}^2 \\ \widehat{\theta^c} \geq \frac{\cos(\theta_{ij}^{(l)}) - \cos(\theta_{ij}^{(u)})}{\left(\theta_{ij}^{(l)} - \theta_{ij}^{(u)}\right)} \left(\theta_{ij} - \theta_{ij}^{(l)}\right) + \cos(\theta_{ij}^{(l)}) \end{cases} \tag{3f}$$

$$\langle \sin(\theta_{ij}) \rangle^S \equiv$$

$$\begin{cases} \widehat{\theta^s} \leq \cos\left(\frac{\theta_{ij}^{(m)}}{2}\right)\left(\theta_{ij} - \frac{\theta_{ij}^{(m)}}{2}\right) + \sin\left(\frac{\theta_{ij}^{(m)}}{2}\right) \\ \widehat{\theta^s} \geq \cos\left(\frac{\theta_{ij}^{(m)}}{2}\right)\left(\theta_{ij} + \frac{\theta_{ij}^{(m)}}{2}\right) - \sin\left(\frac{\theta_{ij}^{(m)}}{2}\right) \\ \widehat{\theta^s} \geq \frac{\sin(\theta_{ij}^{(l)}) - \sin(\theta_{ij}^{(u)})}{\left(\theta_{ij}^{(l)} - \theta_{ij}^{(u)}\right)} \left(\theta_{ij} - \theta_{ij}^{(l)}\right) + \sin(\theta_{ij}^{(l)}) \text{ if } \theta_{ij}^{(l)} \geq 0 \\ \widehat{\theta^s} \leq \frac{\sin(\theta_{ij}^{(l)}) - \sin(\theta_{ij}^{(u)})}{\left(\theta_{ij}^{(l)} - \theta_{ij}^{(u)}\right)} \left(\theta_{ij} - \theta_{ij}^{(l)}\right) + \sin(\theta_{ij}^{(l)}) \text{ if } \theta_{ij}^{(u)} \leq 0 \end{cases} \tag{3g}$$

where $\theta_{ij}^{(m)} = \max\left(\left|\theta_{ij}^{(l)}\right|, \left|\theta_{ij}^{(u)}\right|\right)$

Finally, for (1o), it can be represented with SOCR in the following form [24]:

$$\left|W_{ij}\right|^2 \leq W_{ii} W_{jj} \quad \forall (i,j) \in \mathcal{E} \tag{3h}$$

The specific AC-OPF-QCR model is given as follows, which is also known as the QCR using recursive McCormick envelopes, where the tightness of the convex hull is determined by the variable boundary: the tighter the boundary, the stronger the relaxation [20].

| **AC-OPF-QCR Model** |
|---|
| **minimize:** (1i) |
| **subject to:** (1a), (1c), (1d), (1g), (1h), (1j) - (1m), (2h) |
| $\quad W_{ii} = \langle v_i^2 \rangle^T \quad i \in \mathcal{N} \quad (3i)$ |
| $\quad \Re(W_{ij}) = \langle \langle v_i v_j \rangle^M \langle \cos(\theta_{ij}) \rangle^C \rangle^M \quad \forall (i,j) \in \mathcal{E} \quad (3j)$ |
| $\quad \Im(W_{ij}) = \langle \langle v_i v_j \rangle^M \langle \sin(\theta_{ij}) \rangle^S \rangle^M \quad \forall (i,j) \in \mathcal{E} \quad (3k)$ |

*C. AC OPF Model Based on SDR*

In this section, the AC OPF model based on SDR, abbreviated as AC-OPF-SDR, will be addressed elaborately. By utilizing the newly defined matrix variable $W$ mentioned above, ensuring $W \succcurlyeq 0$ and $\text{rank}(W) = 1$, the complete enforcement of (1o) can be guaranteed [33], as shown in the following expression:

$$W_{ij} = V_i V_j^* \quad \forall i,j \in \mathcal{N} \Leftrightarrow W \succcurlyeq 0 \wedge \text{rank}(W) = 1 \quad (4a)$$

To form the AC-OPF-SDR, the rank constraint is relaxed. The specific AC-OPF-SDR model is given as follows, which is derived from the BIM.

| **AC-OPF-SDR Model** |
|---|
| **minimize:** (1i) |
| **subject to:** (1a), (1c), (1d), (1g), (1h), (1j) - (1m) |
| $\qquad\qquad W \succcurlyeq 0 \qquad\qquad\qquad (4b)$ |

Regarding the following standard SDR form with complex matrix property, abbreviated as AC-OPF-SDR-$\mathbb{C}$:

$$\text{minimize } \text{tr}(A_0 W)$$
$$\text{subject to: } \text{tr}(A_k^{(P,Q)} W) \leq b_k \quad k = 1,2,\ldots,m \quad (4c)$$
$$W \succcurlyeq 0$$

where $\text{tr}(\cdot)$ is the trace operator, and $W$ is a $n \times n$ Hermitian matrix. $A_0 = (Y^H + Y)/2$, where $(\cdot)^H$ denotes the conjugative transposition. Additionally, $A_k^P = (Y^H e_k e_k^T + e_k e_k^T Y)/2$ and $A_k^Q = (Y^H e_k e_k^T - e_k e_k^T Y)/(2\mathbf{i})$, where $Y \in \mathbb{C}^{n \times n}$ represents the admittance matrix, and $e_k$ denotes the $k^\text{th}$ standard basis vector in $\mathbb{R}^n$, respectively.

The AC-OPF-SDR model can be easily transformed into the equivalent form of (3c). Considering the ring isomorphism $\Lambda \coloneqq (\mathbb{C}^{n \times n}) \Rightarrow (\mathbb{R}^{2n \times 2n})$, the (3c) can be further converted to its real-valued form, as shown below:

$$\Lambda(W) \coloneqq \begin{bmatrix} \Re(W) & -\Im(W) \\ \Im(W) & \Re(W) \end{bmatrix} \quad (4d)$$

*Remark 1:*
(a) A complex matrix $W \in \mathbb{C}^{n \times n}$ is positive semidefinite if and only if the real matrix $\Lambda(W)$ is positive semidefinite.
(b) If $W_1, W_2 \in \mathbb{H}^n$, then $\text{tr}(\Lambda(W_1)\Lambda(W_2)) = \text{tr}(\Lambda(W_1 W_2)) = 2\text{tr}(W_1 W_2)$.

By leveraging the properties of ring isomorphism, as given in *Remark 1*, the AC-OPF-SDR model can be converted to a real-valued form with matrix symmetry constraint, referred to as AC-OPF-SDR-$\mathbb{R}$ms model, as given below:

$$\text{minimize } \text{tr}(\Lambda(A_0) X)$$
$$\text{subject to: } \text{tr}(\Lambda(A_k^{(P,Q)}) X) \leq b_k \quad k = 1,2,\ldots,m$$
$$X \succcurlyeq 0, \ X = \begin{bmatrix} B_1 & B_2^T \\ B_2 & B_3 \end{bmatrix} \quad (4e)$$
$$B_1 = B_3, \ B_2^T = -B_2$$

where $X \in \mathbb{S}^{2n \times 2n}$ denotes a $2n \times 2n$ real symmetric matrix, and $X$ is isomorphic to $\mathbb{C}^{n \times n}$.

Additionally, (3f) shows the AC-OPF-SDR model converted from the common SDR [31] without matrix symmetry constraint, and it is referred to as AC-OPF-SDR-$\mathbb{R}$ model, which can be implemented by considering a new real variable $x = \left((\Re(V))^T, (\Im(V))^T\right)^T$.

$$\text{minimize } \text{tr}(\Lambda(A_0) X)$$
$$\text{subject to: } \text{tr}(\Lambda(A_k^{(P,Q)}) X) \leq b_k \quad k = 1,2,\ldots,m \quad (4f)$$
$$X = xx^T \succcurlyeq 0$$

It can be seen from (3f) that the number of matrix elements in (3e) is half of that in (3f).

*D. AC OPF Model Based on CHR*

To implement CHR, it is necessary to simplify these models as given in (3c), (3e), and (3f) obtained from the SDR. At the same time, the Hermitian and positive semidefinite properties of complex and real matrices must be extended to partial matrices.

A clique ($\mathfrak{B} \subseteq \mathcal{N} \times \mathcal{N}$) is a subgraph of an undirected graph such that every two distinct vertices in the clique are connected by an edge. A maximal clique is a clique that cannot be extended by containing an adjacent vertex, which means it is not a subgraph of a larger clique. Besides, the power network $\mathcal{G}$ is chordal graph if every cycle containing 4 or more vertices has a chord. Chordal extension refers to the process of completing the edges $\mathcal{E}$ of $\mathcal{G} = (\mathcal{N}, \mathcal{E})$ to generate a chordal graph $\mathcal{G}^c$.

*Remark 2:*
(a) In the AC-OPF-SDR model, the constraint shown in (3b) is equivalent to $W_{\mathfrak{B}} \succcurlyeq 0$ when applied to the maximal clique in the chordal graph $\mathcal{G}^c$ after performing the chordal extension [24], [32].
(b) For the AC-OPF-SDR-$\mathbb{R}$ model as shown in (3f) and the AC-OPF-SDR-$\mathbb{C}$ model as shown in (3c), since their underlying graphs $\mathcal{G} = (\mathcal{N}, \mathcal{E})$ remain unchanged, the corresponding CHR models are still equivalent to the original models.
(c) Computing clique decomposition on the AC-OPF-SDR-$\mathbb{R}$ and the AC-OPF-SDR-$\mathbb{C}$ models is not equivalent. However, the cost of computing a clique decomposition on the AC-OPF-SDR-$\mathbb{R}$ model is higher [37]. Here, the AC OPF model based on the CHR, abbreviated as AC-OPF-CHR, is given as follows.

| **AC-OPF-CHR Model** |
|---|
| **minimize:** (1i) |
| **subject to:** (1a), (1c), (1d), (1g), (1h), (1j) - (1m) |
| $\qquad\qquad W_{\mathfrak{B}_i} \succcurlyeq 0 \quad i = 1,2,\ldots,k \qquad (4g)$ |

For (4a), a similar approach can be used to formulate the CHR for both complex and real matrices. Here, the CHR models for real matrix and complex matrix variables are referred to as AC-OPF-CHR-$\mathbb{R}$ and AC-OPF-CHR-$\mathbb{C}$, respectively. Additionally, based on the properties of ring isomorphism, the CHR model that converts the complex matrix into the real matrix is referred to as AC-OPF-CHR-$\mathbb{R}$ms.

*Proposition 1:*
val(AC-OPF-CHR-$\mathbb{R}$)=val(AC-OPF-CHR-$\mathbb{R}$ms)=val(AC-OPF-CHR-$\mathbb{C}$)
where "val" denotes the optimal value of the convex relaxation problem.

*Proof:* For the AC-OPF-CHR-$\mathbb{C}$ and AC-OPF-CHR-$\mathbb{R}$ms models, since the AC-OPF-CHR-$\mathbb{R}$ms model is obtained by converting the complex matrix into a real matrix through ring isomorphism, it can be easily proven that the optimal values of the two convex relaxation models are equivalent, that is:
val(AC-OPF-CHR-$\mathbb{R}$ms)=val(AC-OPF-CHR-$\mathbb{C}$) (4h)

Next, the key point of proof is to establish the equivalence between the AC-OPF-CHR-$\mathbb{R}$ms model and the AC-OPF-CHR-$\mathbb{R}$ model. Given that the AC-OPF-CHR-$\mathbb{R}$ms model involves a matrix symmetry constraint, it bears a tighter feasible set, resulting in the following conclusion:

$$\text{val(AC-OPF-CHR-}\mathbb{R}\text{ms)} \geq \text{val(AC-OPF-CHR-}\mathbb{R}\text{)} \quad (4\text{i})$$

Here, it is also necessary to prove that $\text{val(AC-OPF-CHR-}\mathbb{R}\text{ms)} \leq \text{val(AC-OPF-CHR-}\mathbb{R}\text{)}$. First, we will prove that $\text{val(AC-OPF-SDR-}\mathbb{R}\text{ms)} \leq \text{val(AC-OPF-SDR-}\mathbb{R}\text{)}$; then, based on the CHR properties outlined in *Remark 2*, this is equivalent to proving $\text{val(AC-OPF-CHR-}\mathbb{R}\text{ms)} \leq \text{val(AC-OPF-CHR-}\mathbb{R}\text{)}$. By introducing the following definition:

$$\tilde{\Lambda}(X) := (B_1 + B_3)/2 + \mathbf{i}(B_2 - B_2^T)/2 \quad \forall X \in \mathbb{S}^{2n \times 2n} \quad (4\text{j})$$

The inequality $\text{val(AC-OPF-SDR-}\mathbb{R}\text{ms)} \leq \text{val(AC-OPF-SDR-}\mathbb{R}\text{)}$ needs to be proven first. Since $\tilde{\Lambda}(X)$ is a Hermitian matrix, $\Lambda(\tilde{\Lambda}(X))$ satisfies the last two matrix symmetry constraints shown in (3e). Next, it can be demonstrated that $\Lambda(\tilde{\Lambda}(X))$ satisfies $\Lambda(\tilde{\Lambda}(X)) \succcurlyeq 0$. By defining $x = (x_1^T, x_2^T)^T$, we have:

$$\begin{bmatrix} x_1 \\ x_2 \end{bmatrix}^T \Lambda(\tilde{\Lambda}(X)) \begin{bmatrix} x_1 \\ x_2 \end{bmatrix} = \begin{bmatrix} x_1 \\ x_2 \end{bmatrix}^T \begin{bmatrix} B_3 & -B_2 \\ -B_2^T & B_1 \end{bmatrix} \begin{bmatrix} x_1 \\ x_2 \end{bmatrix}$$
$$= \begin{bmatrix} -x_2 \\ x_1 \end{bmatrix}^T \begin{bmatrix} B_1 & B_2^T \\ B_2 & B_3 \end{bmatrix} \begin{bmatrix} -x_2 \\ x_1 \end{bmatrix} \quad (4\text{k})$$

Hence, $\Lambda(\tilde{\Lambda}(X))$ can be expressed as the sum of two positive semidefinite matrices. Finally, it is also necessary to prove that $\text{tr}(\Lambda(A_k^{(P,Q)})X) = \text{tr}(\Lambda(A_k^{(P,Q)})\Lambda(\tilde{\Lambda}(X)))$. Based on the properties outlined in *Remark 1*, it should be noticed that if $A_k^{(P,Q)} \in \mathbb{H}^{n \times n}$ and $X \in \mathbb{S}^{2n \times 2n}$, then:

$$\text{tr}(\Lambda(A_k^{(P,Q)})X)$$
$$= \sum_{i \geq 1, j \leq 2n} \Lambda(A_k^{(P,Q)})_{ij} X_{ji} = \sum_{i \geq 1, j \leq 2n} \Lambda(A_k^{(P,Q)})_{ij} X_{ij}$$
$$= \sum_{i \geq 1, j \leq n} \left[\Re(A_k^{(P,Q)})_{ij}(B_1 + B_3)_{ij} + \Im(A_k^{(P,Q)})_{ij}(B_2 - B_2^T)_{ij}\right]$$
$$= 2\sum_{i \geq 1, j \leq n} \Re\left((A_k^{(P,Q)})_{ij}(\tilde{\Lambda}(X)_{ij})^H\right) \quad (4\text{l})$$
$$= 2\sum_{i \geq 1, j \leq n} (A_k^{(P,Q)})_{ij}(\tilde{\Lambda}(X)_{ij})^H$$
$$= 2\text{tr}(A_k^{(P,Q)}\tilde{\Lambda}(X)) = \text{tr}(\Lambda(A_k^{(P,Q)})\Lambda(\tilde{\Lambda}(X)))$$

Therefore, it is proven that if $X$ is a feasible solution of AC-OPF-SDR-$\mathbb{R}$ model, then $\Lambda(\tilde{\Lambda}(X))$ is a feasible solution of AC-OPF-SDR-$\mathbb{R}$ms model with the same objective value as $X$. After completing the above proof, we have:

$$\text{val(AC-OPF-SDR-}\mathbb{R}\text{ms)} \geq \text{val(AC-OPF-SDR-}\mathbb{R}\text{)}$$
$$\cap \text{ val(AC-OPF-SDR-}\mathbb{R}\text{ms)} \leq \text{val(AC-OPF-SDR-}\mathbb{R}\text{)} \quad (4\text{m})$$
$$\Rightarrow \text{val(AC-OPF-SDR-}\mathbb{R}\text{ms)} = \text{val(AC-OPF-SDR-}\mathbb{R}\text{)}$$

Based on *Remark 2*, the conclusions of the SDR model can be extended to its partial matrices, resulting in the following conclusion:

$$\text{val(AC-OPF-CHR-}\mathbb{R}\text{ms)} = \text{val(AC-OPF-CHR-}\mathbb{R}\text{)} \quad (4\text{n})$$

By combining with (4b), we have:

$$\text{val(AC-OPF-CHR-}\mathbb{C}\text{)} = \text{val(AC-OPF-CHR-}\mathbb{R}\text{ms)}$$
$$= \text{val(AC-OPF-CHR-}\mathbb{R}\text{)} \quad (4\text{o})$$

III. TIGHTENING OF CONVEX RELAXATION

So far, the SDR and QCR techniques have demonstrated their respective advantages in different case studies [11], [19], [26]. Inspired by three existing convex relaxation enhancement techniques [21], such as model intersection, valid inequalities, and boundary tightening, this paper develops a CHR model that can achieve smaller global optimality gap. The rest of this section will discuss how to implement these techniques mentioned above to enhance the CHR.

*A. Model Intersection*

It has been experimentally proven that the recursive McCormick envelopes are difficult to capture the convex hulls of trilinear terms [18]. In [19], a method was proposed to obtain the trilinear or multilinear monomial convex hull using extreme point envelopes, which can yield a tighter relaxation than (2j) and (2k). However, directly constructing the trilinear extreme point envelopes ($\lambda$-based envelopes) will break the potential relationship between the voltage product terms shown in (2b) and (2c). Therefore, in this study, the tighter $\lambda$-based QCR model (referred to as AC-OPF-QC-TLM) proposed in [19] will be adopted, which can effectively capture the convex hull of the trilinear terms while preserving the potential relationship between the voltage product terms. Based on the definition of extreme points [38], we define $\phi(x_1, x_2, x_3) = x_1 x_2 x_3$ with corresponding variable bounds $[\underline{x_1}, \overline{x_1}], [\underline{x_2}, \overline{x_2}], [\underline{x_3}, \overline{x_3}]$. Here, the extreme points of $\phi(\cdot)$ are given by Cartesian product $[\underline{x_1}, \overline{x_1}], [\underline{x_2}, \overline{x_2}], [\underline{x_3}, \overline{x_3}] = \langle \xi_1, \xi_2, \ldots, \xi_k, \ldots, \xi_8 \rangle$. Let $\xi_k^i$ denote the $i^{\text{th}}$ coordinate of $\xi_k$, and the tighter convex envelope of the trilinear term $\phi(x_1, x_2, x_3)$ is given by:

$$\langle x_1 x_2 x_3 \rangle^\lambda = \begin{cases} \hat{x} = \sum_{k=1,2\ldots,8} \lambda_k \phi(\xi_k) \\ x_i = \sum_{k=1,2\ldots,8} \lambda_k \xi_k^i \\ \sum_{k=1,2\ldots,8} \lambda_k = 1, \quad \lambda_k \geq 0, \quad k = 1,2\ldots,8 \end{cases} \quad (5\text{a})$$

For (2b) and (2c), they can be reformulated as follows using (5a):

$$\Re(W_{ij}) = \langle v_i v_j \langle \cos(\theta_{ij}) \rangle^c \rangle^{\lambda_{ij}^c} \quad \forall (i,j) \in \mathcal{E} \quad (5\text{b})$$

$$\Im(W_{ij}) = \langle v_i v_j \langle \sin(\theta_{ij}) \rangle^s \rangle^{\lambda_{ij}^s} \quad \forall (i,j) \in \mathcal{E} \quad (5\text{c})$$

Additionally, by introducing linking constraints as shown in (5d), the constraints between any two variables within the trilinear convex hull are intuitively strengthened.

$$\begin{bmatrix} \lambda_{ij,1}^c + \lambda_{ij,2}^c - \lambda_{ij,1}^s - \lambda_{ij,2}^s \\ \lambda_{ij,3}^c + \lambda_{ij,4}^c - \lambda_{ij,3}^s - \lambda_{ij,4}^s \\ \lambda_{ij,5}^c + \lambda_{ij,6}^c - \lambda_{ij,5}^s - \lambda_{ij,6}^s \\ \lambda_{ij,7}^c + \lambda_{ij,8}^c - \lambda_{ij,7}^s - \lambda_{ij,8}^s \end{bmatrix}^T \begin{bmatrix} v_i^{(l)} \cdot v_j^{(l)} \\ v_i^{(l)} \cdot v_j^{(u)} \\ v_i^{(u)} \cdot v_j^{(l)} \\ v_i^{(u)} \cdot v_j^{(u)} \end{bmatrix} = 0 \quad (5\text{d})$$

By replacing constraints (2i) ~ (2k) in the AC-OPF-QC model with (5b) ~ (5d), the AC-OPF-QC-TLM model can be obtained. Subsequently, by intersecting the AC-OPF-CHR and AC-OPF-QC-TLM models, the obtained relaxation feasible region is strictly smaller than the original feasible region. If the second-order cone constraints shown in (2h) in this model is further replaced by (4a), a tighter $\lambda$-based CHR model with complex matrix, referred to as AC-OPF-CHR-TLM-$\mathbb{C}$ model, can be obtained.

By defining the solution sets of these two models AC-OPF-QC-TLM and AC-OPF-CHR-TLM-$\mathbb{C}$ as $\mathbb{W}_{\text{AC-OPF-QC-TLM}}$ and

$\mathbb{W}_{\text{AC-OPF-CHR-TLM-}\mathbb{C}}$, respectively, we have:

*Proposition 2:* $\mathbb{W}_{\text{AC-OPF-QC-TLM}} \subseteq \mathbb{W}_{\text{AC-OPF-CHR-TLM-}\mathbb{C}}$

*Proof:* For the constraint shown in (4a), every edge $l \in \mathcal{E}$ in $\mathcal{G} = (\mathcal{N}, \mathcal{E})$ is a 2-clique. Therefore, for any edge $l \in \mathfrak{B}$, the 2-by-2 sub-matrices are positive semidefinite, the following conclusion holds:

$$\begin{bmatrix} W_{ii} & W_{ij} \\ W_{ij}^* & W_{jj} \end{bmatrix} \succeq 0 \Leftrightarrow W_{ii}W_{jj} - W_{ij}W_{ij}^* \geq 0 \quad \forall (i,j) \in \mathcal{E}$$
$$\Leftrightarrow |W_{ij}|^2 \leq W_{ii}W_{jj} \quad \forall (i,j) \in \mathcal{E} \tag{5e}$$

which is equivalent to (2h).

*B. Valid Inequalities*

(1) The Lifted Nonlinear Cuts

To further improve the proposed relaxation model, some valid inequalities known as lifted nonlinear cuts (LNCs) [39] can be introduced. These inequalities are valid redundant constraints on the AC-OPF model, which are derived from analyzing the non-convex voltage feasible set, and can be represented by (6a) ~ (6e):

$$v_i^{(\sigma)} = v_i^{(l)} + v_i^{(u)} \quad \forall i \in \mathcal{N} \tag{6a}$$

$$\varphi_{ij} = \left(\theta_{ij}^{(u)} + \theta_{ij}^{(l)}\right)/2 \quad \forall (i,j) \in \mathcal{E} \tag{6b}$$

$$\delta_{ij} = \left(\theta_{ij}^{(u)} - \theta_{ij}^{(l)}\right)/2 \quad \forall (i,j) \in \mathcal{E} \tag{6c}$$

$$v_i^{(\sigma)}v_j^{(\sigma)}\left(\Re(W_{ij})\cos(\varphi_{ij}) + \Im(W_{ij})\sin(\varphi_{ij})\right)$$
$$-v_j^{(u)}\cos(\delta_{ij})v_j^{(\sigma)}W_{ii} - v_i^{(u)}\cos(\delta_{ij})v_i^{(\sigma)}W_{jj} \tag{6d}$$
$$\geq v_i^{(u)}v_j^{(u)}\cos(\delta_{ij}) \times \left(v_i^{(l)}v_j^{(l)} - v_i^{(u)}v_j^{(u)}\right) \quad \forall (i,j) \in \mathcal{E}$$

$$v_i^{(\sigma)}v_j^{(\sigma)}\left(\Re(W_{ij})\cos(\varphi_{ij}) + \Im(W_{ij})\sin(\varphi_{ij})\right)$$
$$-v_j^{(l)}\cos(\delta_{ij})v_j^{(\sigma)}W_{ii} - v_i^{(l)}\cos(\delta_{ij})v_i^{(\sigma)}W_{jj} \tag{6e}$$
$$\geq v_i^{(l)}v_j^{(l)}\cos(\delta_{ij}) \times \left(v_i^{(u)}v_j^{(u)} - v_i^{(l)}v_j^{(l)}\right) \quad \forall (i,j) \in \mathcal{E}$$

The tightness of (6d) and (6e) depends on the boundary parameters of the state variables, which can effectively reduce the size of the feasible region, making the proposed model tighter.

(2) Valid Inequalities Based on Branch Flow Constraints

The effective bounds of branch voltage difference can be determined by the branch thermal limits. In [40], a method was derived to determine the feasible set of the branch angle difference based on the branch thermal limits. On this basis, the valid inequalities for differences of branch voltage magnitudes can be constructed to reduce the feasible set and tighten the model. For convenience of derivation, in this study, the branch is regarded as a $\pi$-model, where the admittance is $g_{ij} + \mathbf{i}b_{ij}$ and the shunt susceptance is $b_{ij}^{\text{sh}}$ for branch $(i,j) \in \mathcal{E}$, then the magnitude of the branch current can then be expressed as follows:

$$I_{ij}(v_i, v_j, \theta_{ij}) = \sqrt{\alpha_1 v_i^2 + \alpha_2 v_j^2 + 2v_i v_j(\alpha_3 \sin\theta_{ij} - \alpha_4 \cos\theta_{ij})}$$
$$I_{ji}(v_i, v_j, \theta_{ij}) = \sqrt{\alpha_2 v_i^2 + \alpha_2 v_j^2 - 2v_i v_j(\alpha_3 \sin\theta_{ij} + \alpha_4 \cos\theta_{ij})} \tag{6f}$$

where $\alpha_1 = g_{ij}^2 + \left(b_{ij} + b_{ij}^{\text{sh}}\right)^2$, $\alpha_2 = g_{ij}^2 + b_{ij}^2$, $\alpha_3 = g_{ij}b_{ij}^{\text{sh}}$ and $\alpha_4 = g_{ij}^2 + b_{ij}^2 + g_{ij}b_{ij}^{\text{sh}}$. Additionally, the corresponding coefficients satisfy the following identity:

$$\alpha_1\alpha_2 \equiv \alpha_3^2 + \alpha_4^2 \tag{6g}$$

Since the maximum current limit is an inequality constraint, the characteristics of this boundary surface can be studied first, expressed as:

$$I_{ij}(v_i, v_j, \theta_{ij}) = I_{ij}^{\max} \quad \forall (i,j) \in \mathcal{E} \tag{6h}$$

Based on (6f), (6g), and (6h), $\theta_{ij}$ can be expressed implicitly by using $v_i$ and $v_j$, and the specific expression is given as follows:

$$\overline{\theta_{ij}} = \arcsin\left(\frac{-\alpha_3\tau \pm |\alpha_4|\sqrt{4\alpha_1\alpha_2 - \tau^2}}{2\alpha_1\alpha_2}\right)$$
$$= \pm\arccos\left(\frac{\tau}{2\sqrt{\alpha_1\alpha_2}}\right) - \arctan\left(\frac{\alpha_3}{\alpha_4}\right) \tag{6i}$$

where $\tau := \alpha_1(v_i/v_j) + \alpha_2(v_j/v_i) - (I_{ij}^{\max})^2/(v_iv_j)$.

Given the physical properties of the branch, for all branches with non-zero $g_{ij}, b_{ij}$, and $b_{ij}^{\text{sh}}$, $\alpha_1$ and $\alpha_2$ are strictly positive. If the argument of the $\arcsin(\cdot)$ function belongs to the real number field and its magnitude does not exceed 1, then (6i) has a real solution, which requires satisfying the following inequality (6j):

$$4\alpha_1\alpha_2 - \tau^2 \geq 0 \Rightarrow -\gamma \leq v_j - \rho v_i \leq \gamma \quad \forall (i,j) \in \mathcal{E} \tag{6j}$$

where $\rho := \sqrt{\alpha_1/\alpha_2}$ and $\gamma := I_{ij}^{\max}/\sqrt{\alpha_2}$.

Therefore, for the box constraints on voltage magnitudes, it can be tighten using constraint (6k) to obtain a feasible region for the voltage magnitudes that is closer to the AC-OPF model [41]. For constraint (6j), the (2d) and (2e) can be leveraged to construct the convex hull of the branch voltage magnitude differences, which requires formulating a set of inequalities to effectively capture the range of possible voltage differences across the branch, as shown in (6k).

$$(v_j - \rho v_i)^2 = \rho^2 W_{ii} + W_{jj} - 2\rho v_i v_j$$
$$= \rho^2 \langle v_i^2 \rangle^T + \langle v_j^2 \rangle^T - 2\rho \langle v_i v_j \rangle^M \quad \forall (i,j) \in \mathcal{E} \tag{6k}$$

Since the constraint (6k) is non-convex, it cannot be directly incorporated into the model. Instead, it can be replaced by an inequality and use convex envelopes, in which the lower and upper bounds $v_{ji}^{(l)}$ and $v_{ji}^{(u)}$ ($v_{ji} \coloneqq v_j - \rho v_i$) are determined by the boundary condition $\gamma$ specified in (6j). Constraint (6l) is the convex relaxation of constraint (6k), which can be used to tighten the feasible set of voltage magnitude differences.

$$(v_j - \rho v_i)^2 \leq \rho^2 \langle v_i^2 \rangle^T + \langle v_j^2 \rangle^T - 2\rho \langle v_i v_j \rangle^M \quad \forall (i,j) \in \mathcal{E}$$
$$v_{ji}^{(l)} \leq v_j - \rho v_i \leq v_{ji}^{(u)} \tag{6l}$$

*C. Bound Tightening*

Based on the properties of the branch thermal limit boundary surface, an algorithm was derived for rapid boundary tightening [40]. By employing (6h) and (6i) and reselecting the coordinate plane, further properties of the branch thermal limit feasible set can be deduced. Hence, $\theta_{ij}^{(u)}$ can be obtained by solving the following two optimization problems:

$$\begin{aligned} & \underset{v_i}{\text{maximize}} & & \overline{\theta_{ij}}(v_i, v_j^{\min}) \\ & \text{subject to} & & v_i^{\min} \leq v_i \leq v_i^{\max} \end{aligned} \tag{7a}$$

$$\begin{aligned} & \underset{v_j}{\text{maximize}} & & \overline{\theta_{ij}}(v_i^{\min}, v_j) \\ & \text{subject to} & & v_j^{\min} \leq v_j \leq v_j^{\max} \end{aligned} \tag{7b}$$

By taking the maximum value of (7a) and (7b) as $\theta_{ij}^{(u)}$, and repeating this process for each branch, the range of $\theta_{ij}$ can be obtained. Compared with the traditional boundary tightening methods, the computational complexity of this method is linearly related to the number of branches in the system. This is because tightening each branch requires constant time, without the need to construct and solve complex constraints. The solutions of (7a) and (7b) can be obtained analytically based on (6i).

In summary, by integrating the aforementioned strategies of model intersection, valid inequalities, and bound tightening, an enhanced SDR model with a tighter feasible set can be constructed, which is referred to as AC-OPF-E-CHR-TLM-$\mathbb{C}$ model , and is given as follows.

---

**AC-OPF-E-CHR-TLM-$\mathbb{C}$ Model**

**minimize:** (1i)
**subject to:** (1a), (1c), (1d), (1g), (1h), (1j) - (1m), (2i), (4a),
(5b) ~ (5d), (6a) ~ (6e), (6l)

---

## IV. AN ACCELERATED SOLUTION ALGORITHM FOR AC-OPF-E-CHR-TLM-$\mathbb{C}$ MODEL BASED ON CLIQUE-GRAPH MERGING STRATEGY

In this section, an accelerated solution algorithm for the developed AC-OPF-E-CHR-TLM-$\mathbb{C}$ model based on clique-graph merging strategy is proposed to achieve efficient solution of OPF problem. By evaluating the size of the SDP and the computation time solved by a kind of commercial interior-point (CIP) solvers, based on a proposed heuristic strategy, the merging of cliques with nephew-uncle or sibling relationships can be effectively achieved, avoiding the selection of hyperparameters. Additionally, by using complex matrices, the number of variables is significantly reduced, greatly enhancing the solution efficiency.

### A. Clique Tree Based Merging Strategy

When computing the SDP problems, various methods can be employed to obtain chordal extensions and maximal cliques. The two most common methods are: (i) Cholesky decomposition with approximate minimum degree (AMD) ordering; and (ii) the minimum degree algorithm. However, the heuristic selection for chordal extension is highly sensitive to the computation time. Regarding the OPF problems, minimizing the number of additional edges is generally inefficient. In view of this, a heuristic clique merging algorithm [27] was proposed, which can effectively enhance the efficiency of solving SDP problems.

*Remark 4:*
(a) The clique intersection property means that for every pair of cliques $\mathcal{B}_i, \mathcal{B}_j \in \mathcal{B}$, their intersection $\mathcal{B}_i \cap \mathcal{B}_j$ is contained within the clique tree connecting $\mathcal{B}_i$ and $\mathcal{B}_j$. The clique union property means that a larger clique $\mathcal{B}_i \cup \mathcal{B}_j$ can replace the original cliques $\mathcal{B}_i$ and $\mathcal{B}_j$, both of which have semidefinite matrix constraints. The increase in dimensionality depends on the size of the overlap between the two cliques.

(b) For a given chordal graph, the Cholesky decomposition with an AMD ordering is usually used to compute the chordal extension, and the Prim's algorithm is used to compute the clique tree [11].
(c) For a given clique, the first clique encountered from the same root node is identified as the parent clique. Conversely, this is referred to as the child clique. If two cliques share the same parent clique, they are considered as siblings. If the parent cliques of two cliques are siblings, one of the child cliques is considered to have an uncle-nephew relationship with the other parent clique.

Taking the IEEE 9-node test system as an example, the corresponding associated graph and clique tree are shown in Fig. 1.

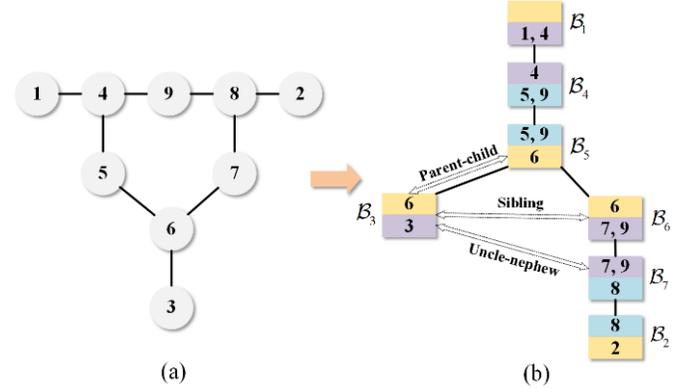

**Fig. 1.** (a) Sparsity graph of IEEE 9-node test system.
(b) Clique tree of IEEE 9-node test system.

As shown in Fig. 1(b), $\mathcal{B}_3$ and $\mathcal{B}_5$ are referred to as having a parent-child relationship; $\mathcal{B}_3$ and $\mathcal{B}_6$ are referred to as having a sibling relationship; $\mathcal{B}_3$ and $\mathcal{B}_7$ are referred to as having an uncle-nephew relationship.

The clique tree based merging strategy usually involves a depth-first order of the parent-child relationship of the clique tree in the graph. When the following two merging criteria are met, the two cliques will be merged heuristically, namely: (i) the evaluation of the size of the SDP; and (ii) the evaluation of the time per iteration of the CIP solver. Additionally, a given threshold $k_{\max}$ is used to determine which evaluation method should be applied for clique merging.

On the one hand, the clique tree based merging strategy makes it difficult to merge cliques with uncle-nephew and sibling relationships. On the other hand, the heuristic strategy requires careful consideration of hyperparameter configuration, such as the threshold $k_{\max}$ and the clique size $L$ used as a convergence criterion, since the solving speed is highly sensitive to these factors. Further, as shown in *Proposition 1* in Section II, the CHR based on real and complex matrices yields the same optimal value. However, owing to the symmetric matrix constraint in AC-OPF-CHR-$\mathbb{R}$ms model, the number of scalars is half that in AC-OPF-CHR-$\mathbb{R}$ model. Therefore, considering CHR in complex form can also improve the computational efficiency.

## B. Clique Graph Based Merging Strategy

To overcome the aforementioned challenges, a clique graph based heuristic merging strategy based on complex matrix SDR is proposed in this paper. The proposed strategy can greedily minimize the computation time of CIP solver and the size of the SDP under complex matrix relaxation by leveraging the structure of the clique graph.

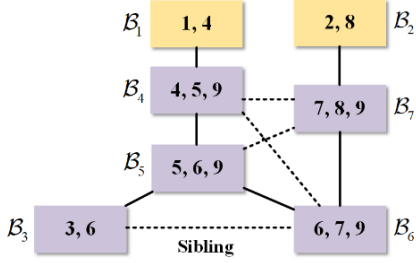

**Fig. 2.** Clique graph of IEEE 9-node test system

The clique graph of the clique tree shown in Fig. 1 (b) is given in Fig. 2. It is worth mentioning that the improved clique graph structure can effectively avoid the situation that cannot be merged in the clique tree structure. The specific implementation process of the proposed heuristic merging strategy is introduced below.

(1) Estimation of CIP Solver Per-Iteration Time

For a given decomposition $\mathfrak{B} = \{\mathfrak{B}_1, \mathfrak{B}_2, \ldots, \mathfrak{B}_i, \ldots \mathfrak{B}_k\}$, let $l_c$ represent the linking constraints between adjacent cliques, and the cardinality of $\mathfrak{B}_i$ is denoted by $|\mathfrak{B}_i|$. Regarding the original primal-dual interior-point method used to solve the SDP problems, it involves the eigenvalue decomposition of the clique matrices and the complexity of the Cholesky factorization for the matrices [30]. In this study, it is assumed that the estimated computation time for the CIP solver is:

$$f_t(|\mathfrak{B}_i|, l_c) = \kappa|\mathfrak{B}_i|^3 + \chi|\mathfrak{B}_i|^2 + \mu(l_c)^3 + \psi \quad (8a)$$

where parameters $\kappa, \chi, \mu$ and $\psi$ are estimated using least squares method.

(2) Estimation of Clique Size

In evaluating the size of the clique, compared with the existing heuristic algorithm [10] used to identify the computational burden, the clique graph based strategy within complex matrix form proposed in this paper can reduce the number of scalars by half, and the function of estimating the computational burden of a SDP problem is given below:

$$f_s(|\mathfrak{B}_i|, l_c) = l_c + |\mathfrak{B}_i|(|\mathfrak{B}_i| + 1)/2 \quad (8b)$$

(3) Algorithm

In this study, the weight $w_{ij}$ of each edge in clique tree is defined as the criterion for heuristic merging, which is determined by $f_t$ and $f_s$:

$$\begin{aligned} &w_{ij}(\mathfrak{B}_i, \mathfrak{B}_j) \\ &= \left(f_t(|\mathfrak{B}_i|, l_c) + f_t(|\mathfrak{B}_j|, l_c) - f_t(|\mathfrak{B}_i| \cup |\mathfrak{B}_j|, l_c)\right) + \\ &\quad \left(f_s(|\mathfrak{B}_i|, l_c) + f_s(|\mathfrak{B}_j|, l_c) - f_s(|\mathfrak{B}_i| \cup |\mathfrak{B}_j|, l_c)\right) \end{aligned} \quad (8c)$$

The edge weight $w_{ij}(\mathfrak{B}_i, \mathfrak{B}_j)$ provides the following criteria for whether the clique merging is reasonable: if $w_{ij} > 0$, the merge is considered beneficial for reducing computation time; otherwise, it is considered to increase the computation time. Using the edge weights $w_{ij}(\mathfrak{B}_i, \mathfrak{B}_j)$, the pseudocode for the clique graph based merging algorithm is given in Algorithm 1:

**Algorithm 1:** The Clique Graph Based Merging Algorithm

**Input:** Cliques in $\mathfrak{B} = \{\mathfrak{B}_1, \mathfrak{B}_2, \ldots, \mathfrak{B}_k\}$
**Output:** Generate a new clique decomposition $\mathfrak{B}_{new}$
1: Generate a weighted clique graph $\mathcal{G}_\mathfrak{B}$
2: **while** $w_{ij}(\mathfrak{B}_i, \mathfrak{B}_j) > 0$ **do**
3:　　Choose $(\mathfrak{B}_i, \mathfrak{B}_j)$ with maximum $w_{ij}$;
4:　　$\mathfrak{B}_{new} \leftarrow (\mathfrak{B} - \{\mathfrak{B}_i, \mathfrak{B}_j\}) \cup \{\mathfrak{B}_i \cup \mathfrak{B}_j\}$
5:　　Update $w_{ij}(\mathfrak{B}_i, \mathfrak{B}_j) \leftarrow \mathfrak{B}_{new}$
6: **end while**

The key idea of Algorithm 1 is to evaluate whether merging two cliques in the clique graph structure will reduce the computation time based on two different evaluation criteria. Additionally, this proposed strategy does not require configuring certain hyperparameters, making it more versatile.

## V. CASE STUDY

In this section, the accuracy and computational efficiency of the proposed AC-OPF-E-CHR-TLM-$\mathbb{C}$ model combined with clique graph based algorithm for solving OPF problems will be assessed elaborately, and compared with the following three recent models and one traditional model:
(1) STCR: Strong tight-and-cheap relaxation.
(2) AC-OPF-QC-TLM: Tighter $\lambda$-based QCR model.
(3) AC-OPF-CHR-$\mathbb{R}$: Real matrix CHR with clique tree based merging strategy on the cases with more than 1000 buses. Otherwise, the merging strategy is not used.
(4) AC-OPF-SDR: Traditional semidefinite relaxation.

All simulations are conducted in MATLAB™ using CVX 2.2, with the solver MOSEK 9.1 (tolerance $\varepsilon = 1.49 \times 10^{-8}$), and the corresponding computer hardware is configured with AMD Ryzen 5 1600X Six-Core Processor. In the simulations, the optimality gap is defined as $100(1 - v_R/\bar{v})$, where $\bar{v}$ is provided by the MATPOWER solver "MIPS", and $v_R$ is given by the convex relaxation model STCR, AC-OPF-QC-TLM, AC-OPF-CHR, SDR and AC-OPF-E-CHR-TLM-$\mathbb{C}$. For some cases, such as 6515rte, "MIPS" fails to find a local optimal solution, so we use IPOPT to find its upper bound. Additionally, the bound tightening algorithms are implemented through a MATLAB package [40].

For the clique merging strategy, the clique graph based merging strategy proposed in this paper is compared with the clique tree based merging strategy proposed in [27]. To evaluate the computational performance advantages, the extreme points constraints in the systems with more than 1000 buses are disregarded. It should be noted that all chordal extensions are based on Cholesky decomposition and AMD ordering. To ensure the fairness in comparison, the optimal $k_{\max}$ mentioned in Section IV is selected to achieve the best computational performance.

Owing to space limitations, the simulation comparisons only provide several representative cases to evaluate the computational accuracy and efficiency of the proposed model, mainly taken from MATPOWER[42] and PGLib-OPF[43].

TABLE I
SIMULATION COMPARISON RESULTS FOR AC POWER FLOW RELAXATIONS (MATPOWER CASES)

| Test case | Optimality gap (%) | | | | |
|---|---|---|---|---|---|
| | STCR | AC-OPF-QC-TLM | AC-OPF-CHR-ℝ | AC-OPF-SDR | Proposed |
| *Small-scale case* | | | | | |
| case5 | 5.22 | 14.54 | 5.22 | 5.22 | 5.15 |
| case57 | 0.00 | 0.06 | 0.00 | 0.00 | 0.00 |
| **Average** | **5.22** | **7.30** | **5.22** | **5.22** | **5.15** |
| *Medium-scale case* | | | | | |
| case118 | 0.02 | 0.25 | 0.00 | 0.00 | 0.00 |
| case300 | 0.01 | 0.15 | 0.00 | 0.00 | 0.00 |
| ACTIVSg500 | 4.20 | 4.42 | 2.11 | 2.11 | 2.09 |
| **Average** | **1.41** | **1.61** | **0.70** | **0.70** | **0.69** |
| *Large-scale case* | | | | | |
| case1354pegase | 0.01 | - | 0.01 | 0.01 | 0.01 |
| case2736sp | 0.01 | - | 0.02 | 0.02 | 0.00 |
| case2746wop | 0.02 | - | 0.01 | 0.01 | 0.01 |
| case3012wp | 0.37 | - | 0.16 | 0.16 | 0.14 |
| case3120sp | 0.12 | - | 0.11 | 0.11 | 0.09 |
| **Average** | **0.11** | **-** | **0.06** | **0.06** | **0.05** |
| *Extra large-scale case* | | | | | |
| case6515rte | 0.16 | - | 0.11 | 0.11 | 0.10 |
| ACTIVSg10k | - | - | 3.75 | - | 0.38 |
| case13659pegase | - | - | 1.30 | - | 1.29 |
| **Average** | **-** | **-** | **1.72** | **-** | **0.59** |

Table I shows the optimality gaps solved by using five different relaxation techniques based on MATPOWER cases. It can be seen that in different sizes of test cases, the proposed AC-OPF-E-CHR-TLM-ℂ model consistently exhibits small optimality gaps.

TABLE II
SIMULATION COMPARISON RESULTS FOR AC POWER FLOW RELAXATIONS (PGLIB-OPF CASE)

| Test case | Optimality gap (%) | | | | |
|---|---|---|---|---|---|
| | STCR | AC-OPF-QC-TLM | AC-OPF-CHR-ℝ | AC-OPF-SDR | Proposed |
| *TYP* | | | | | |
| case3_lmbd | 0.38 | 1.07 | 0.38 | 0.38 | 0.12 |
| case5_pjm | 5.22 | 9.88 | 5.22 | 5.22 | 5.10 |
| case118_ieee | 0.12 | 0.53 | 0.07 | 0.07 | 0.07 |
| case300_ieee | 1.08 | 2.48 | 0.12 | 0.12 | 0.12 |
| **Average** | **1.70** | **3.49** | **1.45** | **1.45** | **1.35** |
| *API* | | | | | |
| case3_lmbd__api | 4.99 | 5.07 | 4.98 | 4.99 | 3.35 |
| case5_pjm__api | 0.20 | 1.74 | 0.20 | 0.20 | 0.05 |
| case118_ieee__api | 22.91 | 21.00 | 10.22 | 10.28 | 6.64 |
| case300_ieee__api | 0.60 | 0.83 | 0.09 | 0.09 | 0.08 |
| **Average** | **7.18** | **7.16** | **3.87** | **3.89** | **2.53** |
| *SAD* | | | | | |
| case3_lmbd__sad | 0.62 | 1.43 | 0.62 | 0.62 | 0.60 |
| case24_ieee_rts__sad | 4.68 | 0.77 | 2.52 | 2.52 | 1.31 |
| case118_ieee__sad | 5.16 | 6.37 | 1.77 | 1.77 | 1.58 |
| case300_ieee__sad | 1.13 | 2.44 | 0.12 | 0.12 | 0.07 |
| **Average** | **2.90** | **2.75** | **1.26** | **1.26** | **0.89** |

Table II shows the optimality gaps solved by using five different relaxation techniques based on PGLib-OPF cases with typical operating condition (TYP), congested operating condition (API), and small angle difference condition (SAD). Similarly, the results indicate that the proposed AC-OPF-E-CHR-TLM-ℂ model can effectively tighten the feasible set. In particular, in the API and SAD cases, the proposed model improves the average optimality gap from 0.37% to 4.65%.

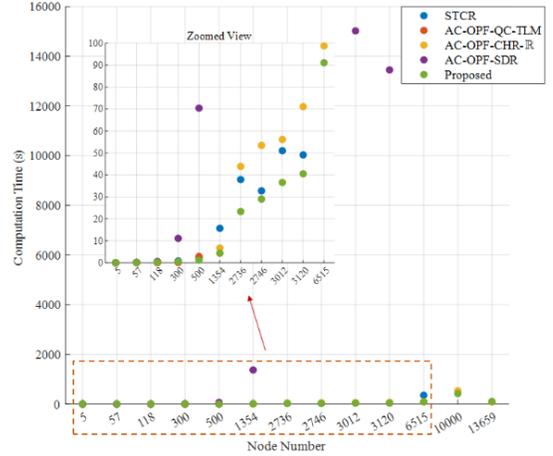

**Fig. 3.** Computation time for the MATPOWER cases

Fig. 3 demonstrates the computational efficiency advantage of AC-OPF-E-CHR-TLM-ℂ model, in which the proposed method improves the computational speed by 7.76% to 46.83% compared with AC-OPF-CHR-ℝ model. It can be seen that performing non-convex relaxation before converting from complex number to real number is beneficial. Additionally, in these cases with more than 1000 buses, the clique graph based merging strategy further improves computational efficiency, with an increase of 1.08 to 1.88 times compared with the original method. Specifically, in the ACTIVSg10k and case13659pegase, the corresponding optimal solutions are obtained in 424.57 seconds and 103.59 seconds, respectively, demonstrating the potential of the proposed model and method for application in extra large-scale systems.

## VI. CONCLUSION

This paper proposes an enhanced SDR model for solving OPF problems, which captures a tighter feasible region by leveraging model intersections, valid inequalities, and boundary tightening techniques. By utilizing ring isomorphism, it can be proven that performing non-convex relaxation before converting from complex number to real number is beneficial. Additionally, by applying branch thermal limits, the valid inequalities for branch voltage differences is derived. In particular, based on the proposed clique graph merging strategy, the model solution is further accelerated on the basis of complex matrix relaxation. Finally, the effectiveness of the proposed model and method is validated on MATPOWER and PGLib-OPF cases. The simulation comparison results show that the proposed model can increase the average optimality gap by 0.01% to 4.65%, while enhancing the computational efficiency by 1.08 to 1.88 times.